\documentclass[preprint,12pt, sort&compress]{elsarticle}

\usepackage{amssymb}
\usepackage{amsmath}
\usepackage{array}
\usepackage{multirow}
\usepackage{xcolor}
\usepackage{float}

\journal{}

\begin{document}

\begin{frontmatter}


\title{Topographic shielding of coastal zones and infrastructure against high tide}



\author[mymainaddress]{P. C. Harisankar}
\author[mymainaddress]{{Tapas Sil}\corref{mycorrespondingauthor}}%
\cortext[Tapas Sil]{Corresponding author}
\ead{tapassil@iiitdm.ac.in}
\affiliation{organization={Department of Physics, Indian Institute of Information Technology, Design and Manufacturing Kancheepuram},
	addressline={Chennai-600127, Tamil Nadu, India}}

\begin{abstract}
High tides are a threat to damage the coast and onshore structures. To investigate mitigation strategies, we simulate waves and a flood-like situation from two-dimensional (2D) dam-break flow with a ramp section at the end of the channel using smoothed particle hydrodynamics (SPH). We analyse the effects of ramps with various topographies to reduce the pressure on structures exerted by the wave. 
Structures of ramp surfaces influence flow behaviour significantly, absorbing kinetic energy of the wave. Increasing the ramp angle reduces the impact on the structure. A wave with a large velocity intensifies the flow impact, rendering the effects on all topography of the ramp almost insignificant. The ramp experiences the highest force exerted by the fluid on the bottom section. 
These insights enhance the understanding of ramp-induced energy dissipation and provide valuable implications for hydraulic engineering and structural resilience.


\end{abstract}




\end{frontmatter}



\section{Introduction}

Events like floods, high tides, tsunamis, and flows from breached dams pose significant threats to onshore structures.  Interaction happens first between the wave and the coast, which reduces the energy of the wave. Roughness of the shore may become critical to determine the intensity of the impact on the onshore object.
Therefore, it is essential to investigate the role of shore-topography in mitigating the impacts of the waves on onshore infrastructures and shoreline erosion.
Our studies will focus on how the angle and surface properties of the slope affect the flow behaviour, including factors such as velocity profiles and pressure distributions.

There are several prescriptions for the remedy 
of reducing the impact of the wave on coastal zone  and infrastructure. As an example, an ecologically designed honeycomb-type revetment reinforced with rigid vegetation is proposed 
which shows a reduction of impact by $40\%$ for overtopping \cite{zhang2022numerical}. Comparison of experimental results of the use of rock bags, polyester netting filled with locally sourced small rocks as a sustainable alternative to traditional rock armour units in coastal revetments \cite{moss2025innovative} reveals that rock bags reduce the wave by an average of $53\%$ compared to the rock armour model. In some cases, the grass blocks are considered to mitigate the coastal damage \cite{barendse2022hydrodynamic}. Seawalls and submerged breakwaters are used to protect onshore objects
\cite{van2024numerical}.
The flow mitigation is also achieved on the flat shore using various types of obstacles, including a single piece with a trapezoidal shape and flat front surfaces, or an arch of multiple angles, which have been found to effectively reduce the maximum pressure on the dam surface.
An arched obstacle achieved even greater reductions in maximum pressure \cite{issakhov2018numerical}.  
The wave overtopping on different sea wall structures, i.e. stepped-face wall, curved, inclined and vertical walls, is studied \cite{dang2021numerical}. Overtopping water was found to be reduced most significantly in the case of a stepped-face wall. Impact of a flow of water over a submerged plain ramp on the onshore wall obtained from SPH calculations was compared against experimental data \cite{rostami2015numerical}.  Dam-break flows interacting with offshore trapezoidal and triangular obstructions, focusing on the effects of slanted-shaped objects in obstructing the flow \cite{ozmen2011,simsek20232d}.

The dam-break model, where the sudden release of a water column under gravity generates waves, may be considered to be one of the wave makers, which mimics the sudden flood and high tide situations \cite{yang2018numerical,tian2021smooth,simsek20232d,Maranzoni2023,Aureli2024,peramuna2024review}. 
Three methods are commonly employed for studying such problems numerically: the fixed grid method, the moving grid method, and the mesh-free method.
Fixed grid methods, such as the volume of fluid (VOF) \cite{ozmen2011,issakhov2018numerical} method, or Lattice Boltzmann method (LBM) \cite{purbasari2018,watanabe2021}, utilise a stationary grid to track fluid movement and interface interactions.
Moving grid methods, such as the finite element method (FEM) \cite{zhang2017simulation,zhang2018evaluation}, dynamically adjust the grid to capture changes in geometry or flow features. 
Mesh-free methods, such as smoothed particle hydrodynamics (SPH) \cite{cherfils2012josephine,marrone2011delta}, operate without a defined grid structure and directly track particles.

Previous studies have not explored the effect of modifying slanted surfaces from smooth to hybrid shapes to obstruct the flow. In this article, we examine dam-break flow scenarios involving a plain and a modified slanted slab, followed by a flat shore, to analyse the wave impact on the onshore wall. We chose SPH for our studies because it has been found to be highly effective for modelling significant deformation problems compared to grid-based simulation methods \cite{harisankar2023drop,harisankar2025zigzag,sil2023nanofluid}. This article provides a brief overview of the SPH formulation, followed by the results and discussion section. The study begins with model validation, proceeds with a description of the model and analysis of the results, and concludes with a summary of key findings. 
\section{\textbf{Formulation}}\label{sec2}
SPH is a mesh-free Lagrangian simulation method developed in 1977 \cite{gingold1977,lucy1977} to study astrophysical phenomena. Over time, the method has been extended to a wide range of applications in science and engineering. SPH has become a popular tool among researchers for simulating fluid dynamics because of its ability to handle free surface tracking and moving boundary problems.

SPH method consists of two approximation methods: kernel approximation and particle approximation. Firstly, the Kernel approximation represents a function and its derivatives in continuous form as an integral representation. The kernel approximation of a function and its derivatives are \cite{gingold1977,liu2003},
\begin{equation}
	\langle f(\mathbf{x})\rangle = \int_\Omega f(\mathbf{x}^\prime)W(\mathbf{x} - \mathbf{x}^\prime,h)\mathbf{dx}^\prime,
	\label{eq:funAvg}
\end{equation} 
\begin{equation}
	\langle\nabla f(\mathbf{x})\rangle = -\int_\Omega f(\mathbf{x}^\prime)\nabla W(\mathbf{x} - \mathbf{x}^\prime)dx^\prime,
	\label{eq:GradfunAvg}
\end{equation} 
where, $W(x - x^\prime,h)$ is the kernel function, which has delta function-like properties. The kernel function is defined considering neighbouring particles within a distance $h$, known as the smoothing length. 
The particle approximation represents the problem domain using a set of particles and estimates the observables, such as velocity, density, stress, etc., for this set of particles. The particle approximation for a function $f(x)$ and its gradient for the particle at  $a$ is given by:
\begin{equation}
	\langle f(\mathbf{x}_a)\rangle =\sum_b \frac{m_b }{\rho_b}f(\mathbf{x}_b)W_{ab},
	\label{eq:funSPH}
\end{equation}
\begin{equation}
	\langle \nabla f(\mathbf{x}_a)\rangle =\frac{1}{\rho_a}\left[\sum_b m_b(f(\mathbf{x}_b)-f(\mathbf{x}_a))\nabla W_{ab}\right],
	\label{eq:funAvgSPH}
\end{equation}
where the suffix $a$ and $b$ represent two SPH particles. In Eq. \ref{eq:funSPH}, $\rho$ represents density, $m$ is the mass, and $W$ is the kernel function. The B-spline kernel function based on the cubic spline \cite{monaghan1985} is used here, which is written as
\begin{equation}
	W_{ab}=\alpha_d%
	\begin{cases}
		\frac{2}{3}-s^2+\frac{1}{2}s^3, & 0\le s <1\\
		\frac{1}{6}(2-s)^3, & 1\le s <2\\
		0, &  s \ge 2,\\
	\end{cases}
	\label{eq:Kernal}
\end{equation}
where for 2D,
\begin{equation}
	\alpha_d=\frac{15}{7\pi h^2} \:\: \text{and} \:\: 	s=\frac{|\bold{x}_a-\bold{x}_b|}{h}. 
	\label{eq:Kerneldim}
\end{equation}
Here, 	$\mathbf{x}_a$ and $\mathbf{x}_b$ are the position vectors of particle $a$ and particle $b$. 
The Navier-Stokes equation for conservation of massand momentum  is given as, \cite{adami2012generalized}.
\begin{equation}\label{eq:continuity}
	\frac{D\mathbf{\rho}}{Dt}=-\rho\nabla.\mathbf{u},
\end{equation}
\begin{equation}\label{eq:momentum}
	\frac{D\mathbf{u}}{Dt}=-\frac{1}{\rho}\mathbf{{\nabla}}P+\frac{\mu}{\rho}\nabla^2 \mathbf{u}+\mathbf{g},
\end{equation}
where, $\mathbf{u}$ is the velocity, $P$ is pressure, $\mu$ is viscosity and $g=(0,0,-9.8m/s)$ acceleration due to body force $\mathbf{F_b}$. The pressure is calculated using Tait's equation of state,
\begin{equation}
	P = \frac{c_{0}^2\rho_0}{\gamma}\left\lbrace
	\left(\frac{\rho}{\rho_0}\right)^{\gamma} -1\right\rbrace,
	\label{eq:TaitEOS}
\end{equation}
where $c_0=10u_{max} (u_{max} =\sqrt{2gH})$ is the speed of sound, $\rho_0$ is the reference density, and $\gamma$ is a polytropic constant which varies from 1 to 7, for an example, for water $\gamma=7$.
The SPH discretised form of Eq. (\ref{eq:continuity}) and Eq. (\ref{eq:momentum}) is given as \cite{hu2006multi,adami2012generalized}, 
\begin{equation}
	\frac{D\mathbf{\rho_{a}}}{Dt}=\rho_a\sum_b \frac{m_b}{\rho_a}(\mathbf{u}_a-\mathbf{u}_b).\mathbf{\nabla W_{ab}},
	\label{eq:ContSPH}
	\underline{}
\end{equation}

\begin{eqnarray}
	\begin{split}
		\frac{D\mathbf{u}_{a}}{Dt}=&-\frac{1}{m_a}\sum_b \left(V_{a}^{2}+V_{b}^{2}\right)\tilde P_{ab}\mathbf{\nabla W_{ab}}\\
		&+\frac{1}{m_a}\sum_b \left(V_{a}^{2}+V_{b}^{2}\right)\tilde \mu_{ab}\frac{\mathbf{x}_{ab}.\mathbf{\nabla W_{ab}}}{|\mathbf{x}_{ab}|^{2}+\epsilon h^2}\left( \mathbf{u}_a-\mathbf{u}_b\right)+\mathbf{g},\\		
		\label{eq:MomSPH}
	\end{split}
\end{eqnarray}
where,
\begin{equation}
	\tilde P_{ab}=\frac{P_a\rho_b+P_b\rho_a}{\rho_a+\rho_b},
\end{equation}
\begin{equation}
	\tilde \mu_{ab}=\frac{\mu_a\mu_b}{\mu_a+\mu_b},
\end{equation}
$V_{i}$ is the volume SPH particle, $\mathbf{x}_{ab}=\mathbf{x}_{a}-\mathbf{x}_{b}$.
A XSPH formula is adapted for stable fluid flow. The updated velocity field after XSPH correction is given as \cite{monaghan1989xsph}, 
\begin{eqnarray}
	\frac{d\mathbf{r}_{a}}{dt}=\mathbf{u}_{a}+
	\epsilon\sum_{b}m_{b}\frac{\mathbf{u}_{ab}}{\bar{\rho}_{ab}}W_{ab}
	\label{eq:XSPH}
\end{eqnarray}
The fluid-solid interaction is considered \cite{adami2012generalized,song2018} by iterating the fluid velocity and pressure to the dummy wall particles,
\begin{equation}
	\mathbf{u}=2\mathbf{u}_w-\hat{\mathbf{u}},
\end{equation}
where $\mathbf{u}_w$ is the real wall velocity and $\hat{\mathbf{u}}$ is,
\begin{equation}
	\hat{\mathbf{u}}=\frac{\sum \mathbf{u}_{wf}W_{wf}}{\sum W_{wf}},
\end{equation}
Similarly wall pressure is calculated as,
\begin{equation}
	P=\frac{\sum P_{wf}W_{wf}}{\sum W_{wf}},
\end{equation}
 We have developed a SPH code using FORTRAN to solve the aforementioned equations subjected to the appropriate boundary conditions.
\section{Validation of results}
We validate our FORTRAN code for SPH modelling by simulating a two-dimensional (2D) dam-break flow for non-dimensional (NDML) \cite{adami2012generalized} and dimensional (DML) cases  \cite{chen2015sph} with the inputs as given in the respective references mentioned above, and by comparing our results with available experimental data. 
\begin{figure}
	\centering
	\includegraphics[width=0.7\textheight]{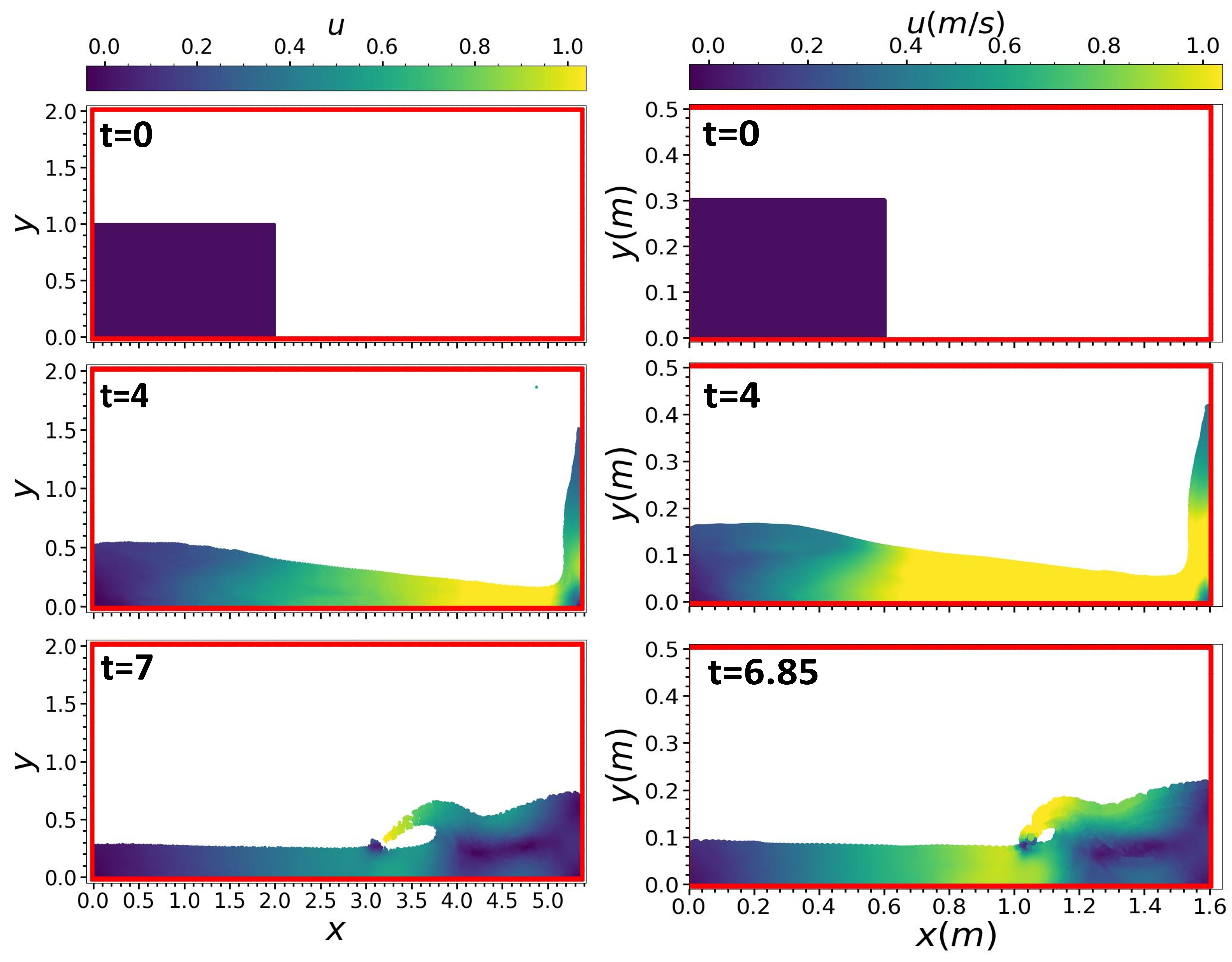}
	\caption{Velocity contour from  dam-break simulations. Figures in the left column represent non-dimensional dam-break model discussed in \cite{adami2012generalized}, whereas those in the panels of the right column, show the dimensional dam-break simulation \cite{chen2015sph} }.
	\label{Fig:damValDyn}
\end{figure}
The snapshots of velocity contours presented in  Fig. \ref{Fig:damValDyn} show that the simulated wave profiles mimic well with earlier theoretical and experimental results both for the  NDML and DM  \cite{adami2012generalized,chen2015sph}.
We plot pressure at $y= 0.19$ (and $y=0.06$m) on the opposite wall ($P_{y}$) versus time ($t$) graph and compare the results obtained from our code with the experimental data \cite{zhou1999} as shown in Fig. \ref{Fig:ValiPres} (NDML on in the left panel $a$ and DML is in on the right panel $b$). Both NDML and DML graphs generated by our code are in good agreement with experimental data as well as with the previous SPH simulations \cite{adami2012generalized,chen2015sph}.

\begin{figure}
	\centering
	\includegraphics[width=0.7\textheight]{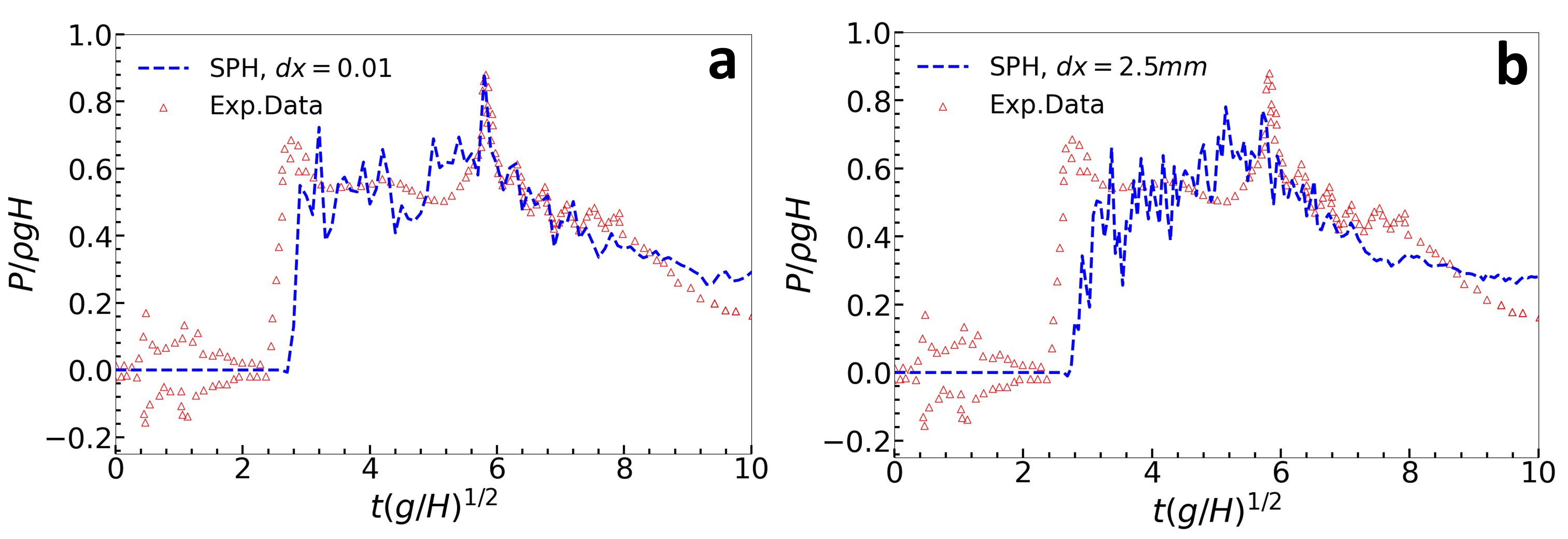}
	\caption{Variation of pressure  on the wall with time to results obtained from our SPH calculation with parameters considered in \cite{adami2012generalized} (panel $a$) and with parameters in  \cite{chen2015sph} (panel $b$). The results are compared with experimental data \cite{zhou1999}.}
	\label{Fig:ValiPres}
\end{figure}
\begin{figure}
	\centering
	\includegraphics[width=0.9\textwidth,height=0.4\textheight]{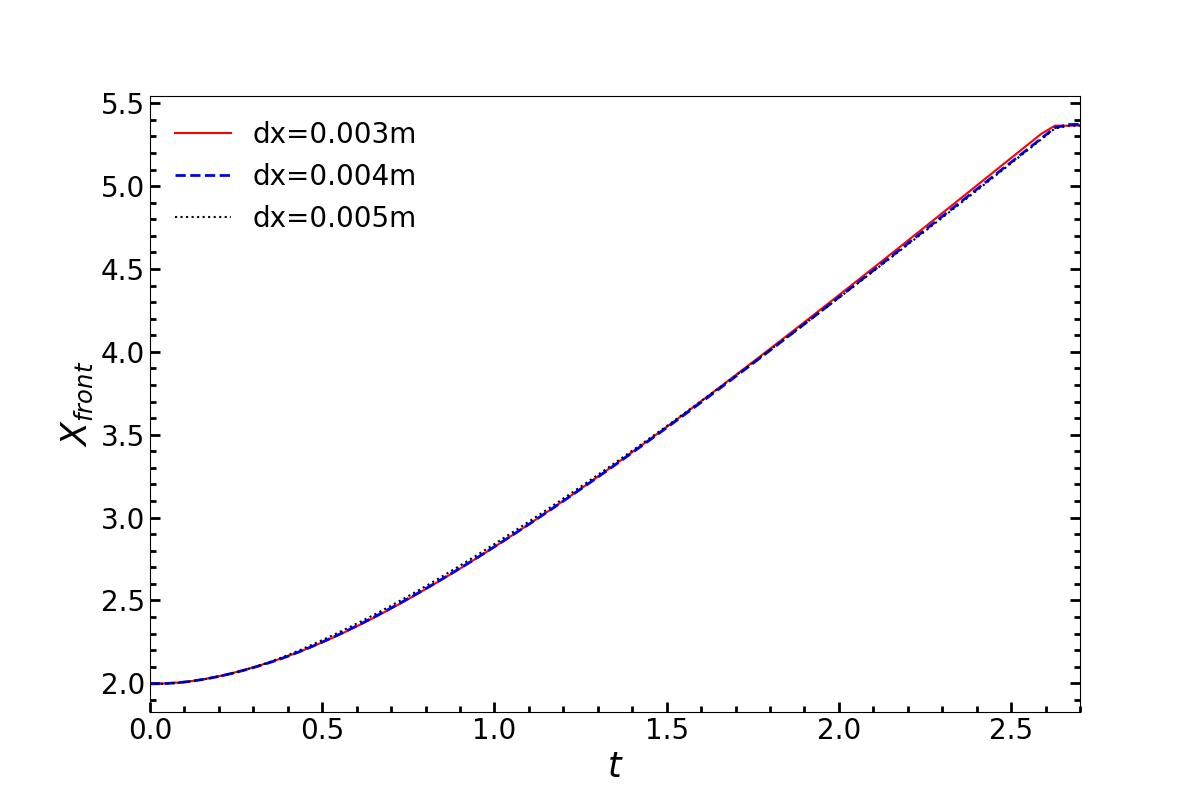}
	\caption{Bore tip  position with time for different mesh size \cite{mokrani2016} (various particle number $N$ ).}
	\label{Fig:conv}
	\end{figure}
\begin{figure}
	\centering
	\includegraphics[width=0.7\textheight]{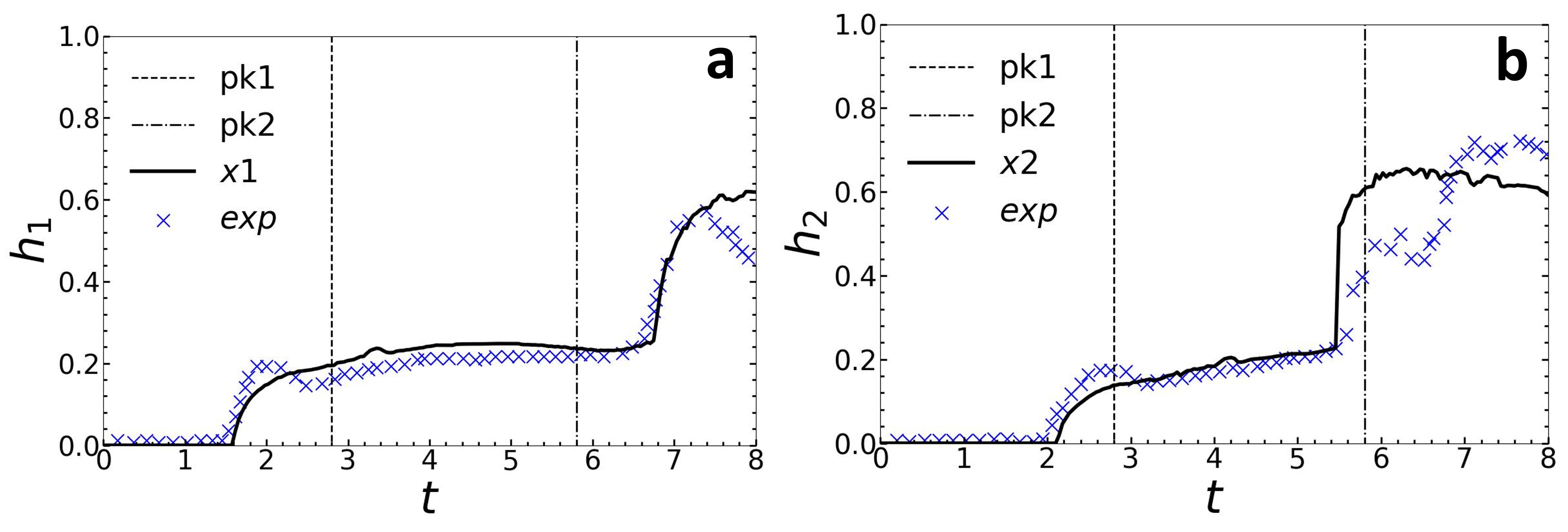}
	\caption{Time evolution of water surface elevation at (a) $x_1=3.7$ and (b) $x_2=4.54$   compared with experiment of \cite{zhou1999}.}
	\label{Fig:WaterLevel}
\end{figure}

A way of testing convergence of the results of a code against SPH particle numbers (or initial spacing between particles $dx=dy$) is to check that the trajectory of the bore front (the bottom-most point of the dam and which is nearest to the opposite wall) \cite{mokrani2016}.
We consider the breaking of  a 2D dam ($2\times 1$ units, NDML). The water wave hitting a wall at $L$ ($=x=5.36$) for particle spacing  $dx = 0.003, 0.004, \text{ and } 0.005$ at the initial stage. We track the bore front motion until it impacts the wall and are plotted ($x$ versus $t$) in Fig. \ref{Fig:conv}. It is noted that the trajectories remain almost coincident with each other, which, for the $dx$s considered here, confirms the convergence. We have taken $dx = 0.005 m$ for all calculations presented in this article.
We examined the water level at two points, $x_1=1.0$m and $x_2=0.495$m, as shown in Fig. \ref{Fig:WaterLevel} and compared with the experimental results \cite{mokrani2016}. At $x_1$, the numerical results closely follow experimental observations over a long period of time (up to $t=7.3$s). Beyond this $t$ (time), the theoretical water surfaces are slightly higher than the experimental water surface. At $x_2$, the agreement with experimental results is excellent up to $t=5.8$s, beyond which the theoretical curve goes over the experimental curve and then comes down after crossing the experimental curve. The water-level curves obtained from simulations with our code agree satisfactorily with the experimental curves.
\section{Results and discussion}
\begin{figure}
	\centering
	\includegraphics[width=0.7\textheight]{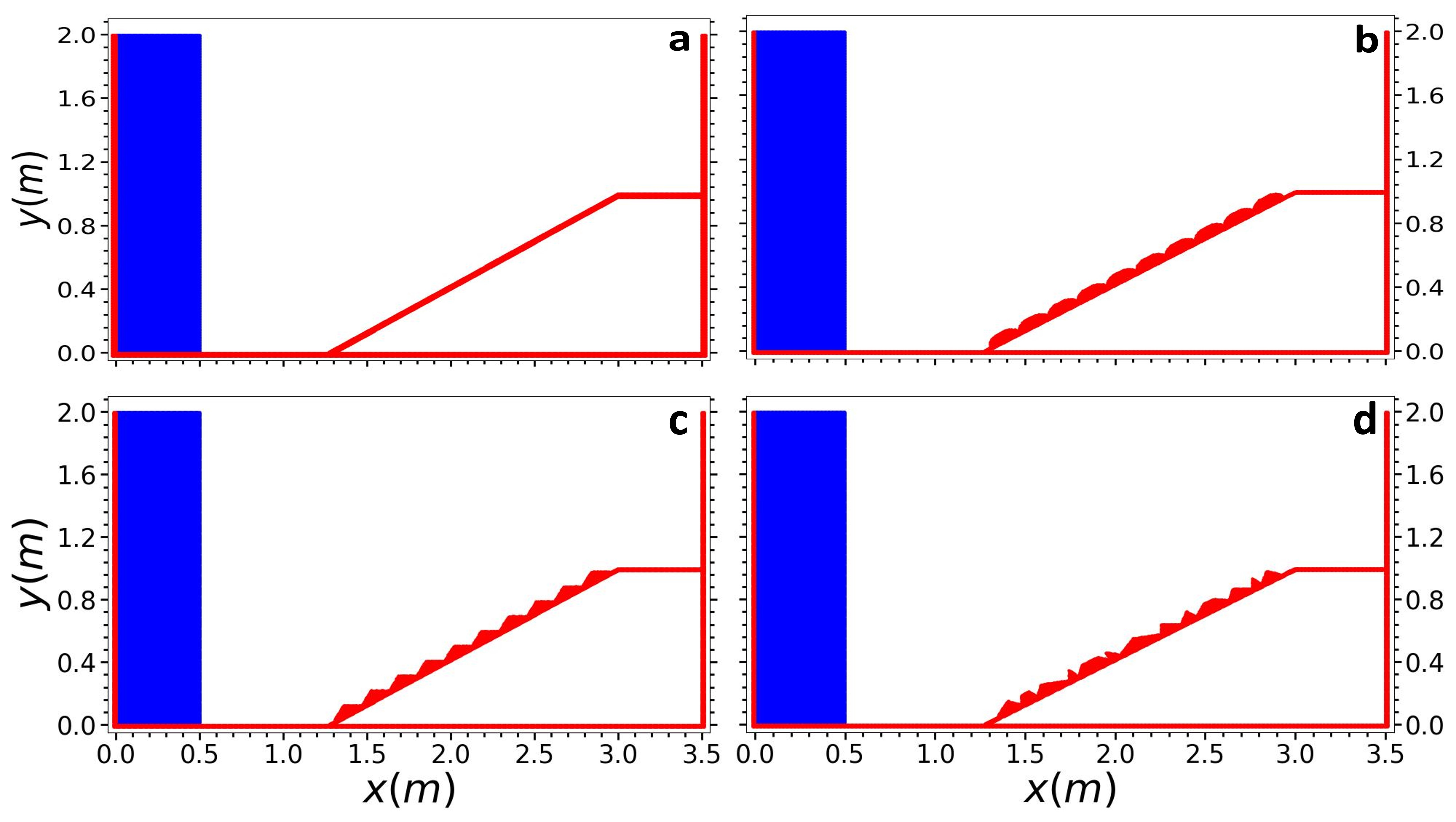}
	\caption{Initial model ($t=0s$) for a dam height of $2.0$m. Panel (a) represents a plain ramp, (b) semicircle(SC) (c) triangular(TR) and (d)  represents randomly shaped ramp(RA).}
	\label{Fig:ini}
\end{figure}
To investigate the role of the structure of the coastal surface in reducing the impact on the onshore object, we consider coastal surfaces with a slanting angle $\theta$ (ramp) which meets an onshore horizontal platform (HP) at a height of $1$m from the bottom of the water body. We consider a structure (e.g., a wall) located at $0.5$m from the coastline. 
A water reservoir of length $L_x = 0.5$m and height $H$m is placed on the other end of the ramp and wall system within the workspace of length $L=3.5$m, which breaks down under gravity and generates a surge of water flow towards the ramp.
As a sample of the initial condition of the system for simulations, we present in Fig. \ref{Fig:ini} corresponding to $H = 2.0m$,  $\theta=30^o$  of various ramp topographies, i.e., ramp surface configurations (RSC): plain (PL), semicircular (SC), triangular (TR), and random-shaped (RA). Each structure has an amplitude (height from the base) of $0.05$m. 
\begin{table}[ht]
	\centering
	\caption{Maximum height $h^a$ (in m) of water reached along the wall and the corresponding time (in s) ginen in the parenthesis for $\theta=30^0$ ramp.} 
	\begin{tabular}{|c|c|c|c|c|}
		\hline
		$H$ & $h^a_{PL}$ ($t$) & $h^a_{SC}$ ($t$) & $h^a_{TR}$ ($t$) & $h^a_{RA}$ ($t$)\\ 
		\hline
		1.0 & 1.25 (1.30) & 0.65 (1.04) & 0.75 (0.93)& 0.80 (0.91) \\
		\hline
		1.5 & 1.75 (1.28) & 0.95 (1.20)&1.00 (1.07)& 1.05 (1.00)\\
		\hline
		2.0  & 2.45 (1.40) & 1.60 (1.42)&1.55 (1.38)&1.47 (1.36)\\
			\hline
		2.5  & 2.70 (1.49) & 1.95 (1.43)&1.85 (1.44)&1.85 (1.46)\\
		\hline
		3.0 & 2.85 (1.63) & 2.65 (1.53)&2.60 (1.51)&2.60 (1.53) \\
		\hline
	\end{tabular}
	\label{tab:1}
\end{table}

Table \ref{tab:1} displays the height ($h^a$) ascended by the wave water along the ramp plus wall and time $t$ to reach $h^a$ for various dam heights $H$. The slope of the ramp is $\theta=30^o$. Ascending height increases with an increase in $H$, which is evident as kinetic energy is larger for a higher $H$ that pushes the water to a higher $h^a$. Each column in the table represents the results of different RSC as shown in Fig. \ref{Fig:ini}. For $H=1.0$m and $1.5$m, water does not possess sufficient energy to climb up the ramp and hit the wall ($h^a\le 1.0$m), except for the plain ramp. For $h=2.0$m, $h^a$ is minimum for RA ($1.47$m, about a $40\%$ reduction compared to PL), indicating that the energy-dissipating ability is greater in RA than in SC and TR. For $H=3.0$m, the reduction of $h^a$ from PL is approximately ($9.0\%$) similar for all configurations, which means that at high wave velocities, the revetment geometry of the ramp becomes less effective. We may need to increase the amplitude of the ramp's modifying structures. 

\begin{figure}
	\centering
	\includegraphics[width=0.7\textheight]{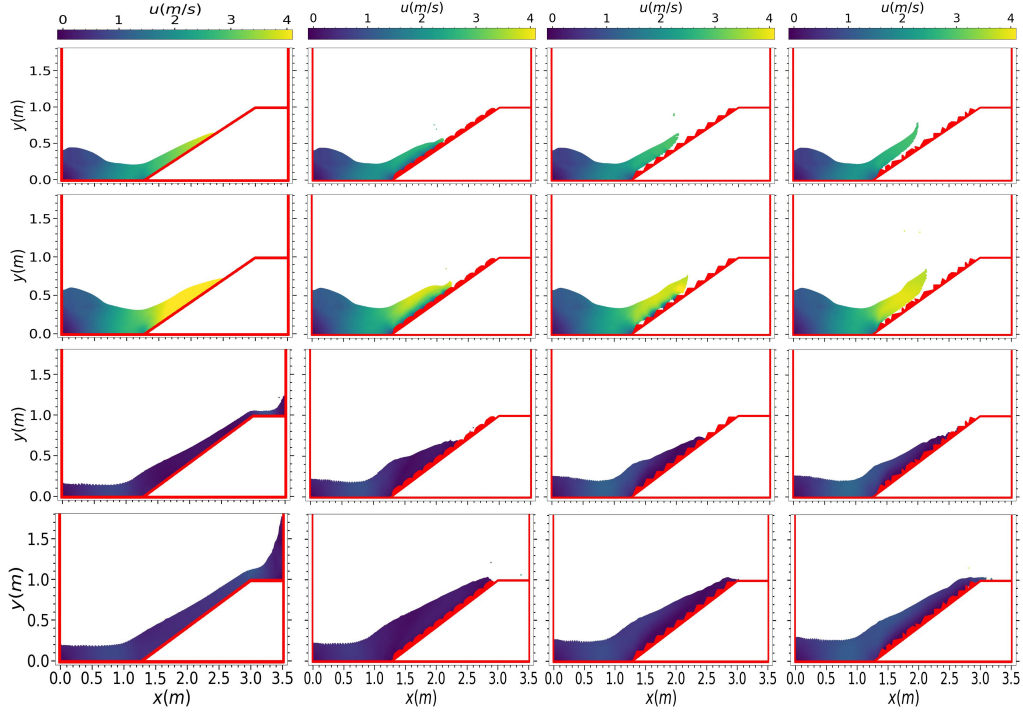}
	\caption{Velocity contour of the flow from dam heights $H=1.0$m and $1.5$m ($\theta=30^o$) at $t=0.6$s ($1^{st}$ and $3^{rd}$ columns) and at the time of maximum  ascent ($h^a$) of water ($2^{nd}$ and $4^{th}$ columns) along the ramp and onshore structure (a wall) for different surfaces of the ramp.}
	\label{Fig:VelContH11o0to1o5}
\end{figure}

\begin{figure}
	\centering	
	\includegraphics[width=0.7\textheight]{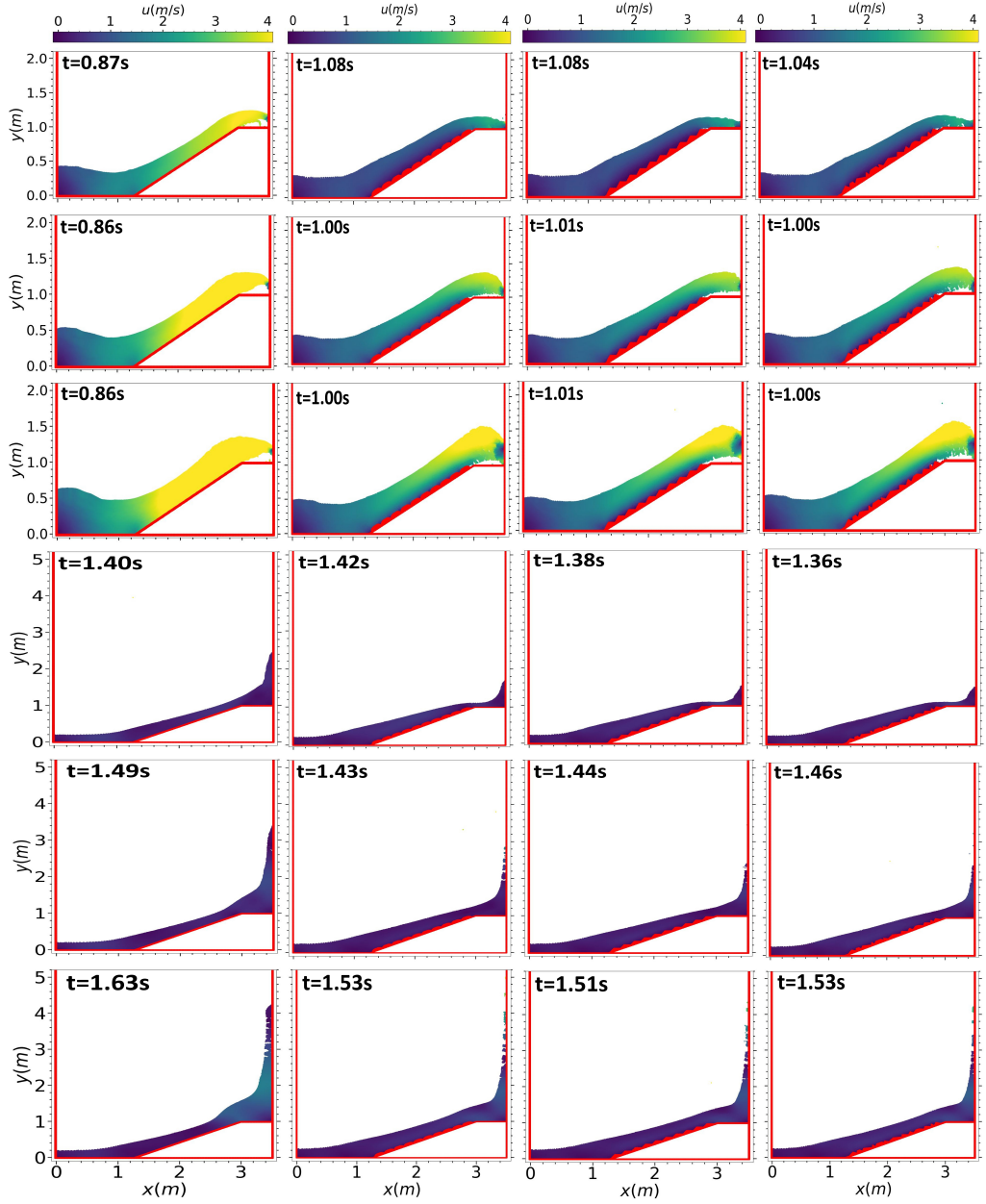}
	\caption{Initial impact on wall (Rows 1-3) and Maximum flow height (Rows 4-6) achieved for H=$2.0$m (row-1,4),$2.5$m (row-2,5) and $3.0$m (row-3,6), for $\theta=30^\circ$ for different ramp structure. 
	Snapshots at the time impact on the wall for $H=2.0$m $2.5$m and $3.0$m  ($\theta=30^o$) in first, second and third columns. The fourth colomn respresents maximum height water travelled for $H=2.0$m and $3.0$m.}
	\label{Fig:dynaH2-3 30deg}
\end{figure}

Figure \ref{Fig:VelContH11o0to1o5} illustrates the velocity contour of the flow from dams of heights $H=1.0$m and $1.5$m during the ascent of water along the ramp ($\theta=30^o$) and coastal structure (wall). Each column in the figure represents the result for ramps of different topographies. As seen from this figure as well as from the table \ref{tab:1}, the maximum ascending height is achieved for the ramp with PL. For other RSCs, the maximum ascending heights are as follows: $h^a_{SC}< h^a_{TR}< h^a_{RA}$, for $H=1.0$m and $1.5$m, which we expect to have in a reversed order. The reason for the above anomaly is that water slides up the SC, which are smoother than the TR and RA. For TR and RA, the water jumps as a jet and lands at a distance slightly farther from the expected one  (first two rows; plotted at $t=0.6$s). The order of the $h^a$ values reverses for $H=2.0$m and above. 
Figure \ref{Fig:dynaH2-3 30deg} shows the velocity contour at the time of wave impacts on the wall (just before water spreads on the wall) for $H=2.0$m, $2.5$m and $3.0$m ($\theta=30^o$) in the first, second, and third rows, respectively. Rows four to six display the velocity contour at the time of maximum $h^a$ for $H$ mentioned above. These diagrams corroborate the results presented in Table \ref{tab:1}. For $H=2.0$m water slide on the ramp and horizontal platform before hitting the bottom ($y=1.05$m) of the wall for all RSCs except PL. Whereas, for $H=2$m and PL, a water jet directly hits the wall at a point slightly above ($y=1.2$m) the base of the wall without touching the horizontal platform. The same way the water jets hit the wall directly for $H=2.5$m and $3.0$m for all RSC. These plots will help us understand the maximum pressure ($P^m$) and its location on the wall in the later part of this article. 

\begin{figure}[ht]
	\centering
	\includegraphics[width=0.7\textheight]{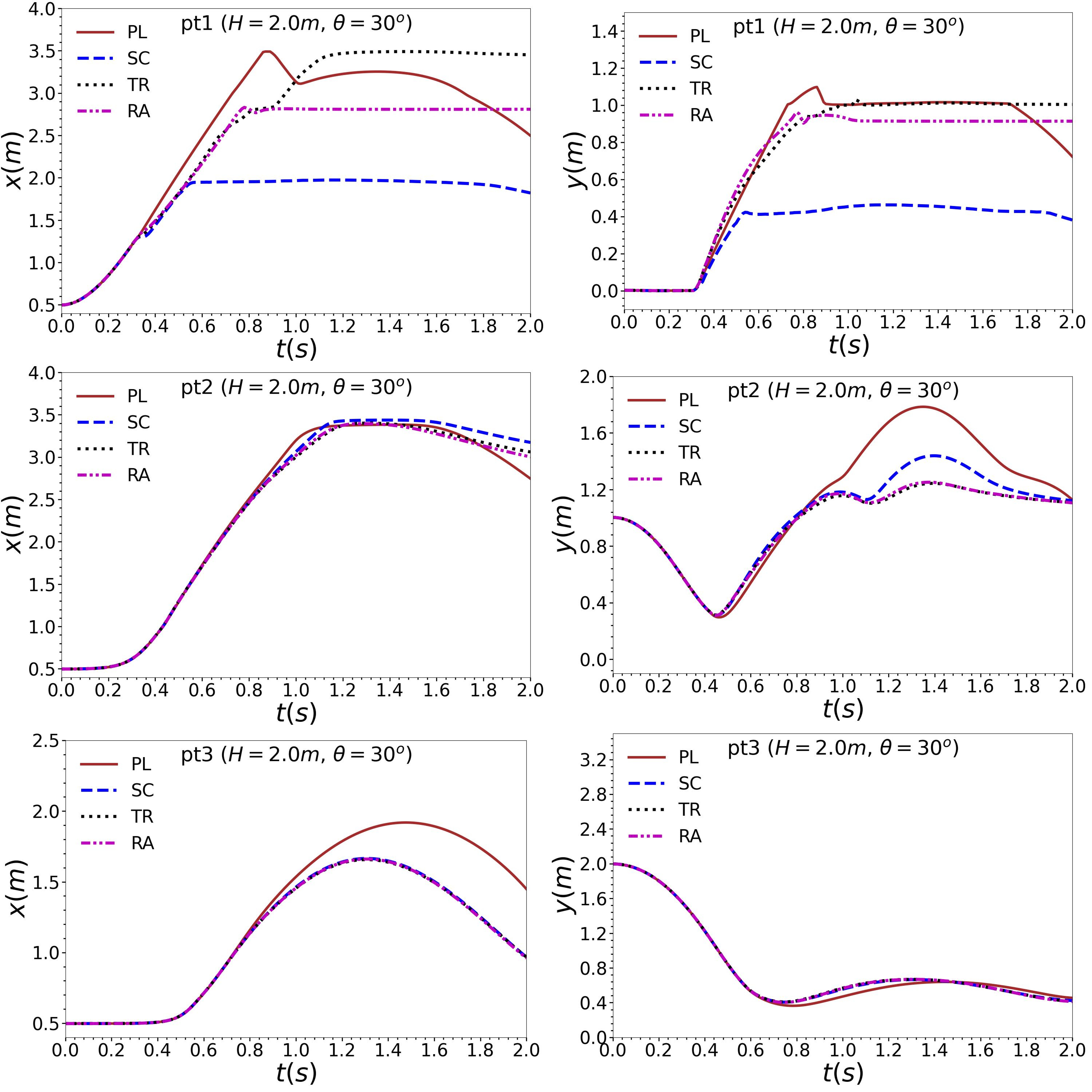}
	\caption{Motion of SPH fluid particles at initial point pt1, pt2, pt3 for $H=2.0m$ and  $\theta=30^o$}
	\label{Fig:track}
\end{figure}

The flow behaviour is analysed by tracking fluid particles at three points of the initial fluid column. In Fig. \ref{Fig:track}, the $x$-axis and $y$-axis movements of a fluid particle are plotted for all ramp shapes at $\theta=30^o$ with $H=2.0$m, considering points pt1 ($0.5$m, $0.0025$m($=dx/2$)), pt2 $(0.5,1.0)$, and pt3 $(0.5,2.0)$. Tracking the $x$-axis motion of the particle reveals that pt3, which is at the top of the water column, is minimally affected by the ramp structure of the ramp surface. The points at the bottom and the middle ascend the ramp and impact the vertical wall, whereas pt3 climbs halfway up the ramp and then descends. Examining the y-axis movement of the particle displayed in Fig. \ref{Fig:track}, pt1 climbs up the ramp and settles in the horizontal slab region.
Meanwhile, in the plain ramp case, pt2 reaches its maximum height before descending. Modifications to the ramp topography with triangular and random shapes significantly reduce the momentum of pt2, resulting in the lowest peak height in the displacement ($y$) vs time graph. Conversely, the semicircular ramp allows the fluid to gain slightly more momentum compared to the RA and TR ramps. The fluid particle at pt3 reaches the middle of the slanted slab before beginning its descent.
\begin{figure}[ht]
	\centering
	\includegraphics[width=0.7\textheight]{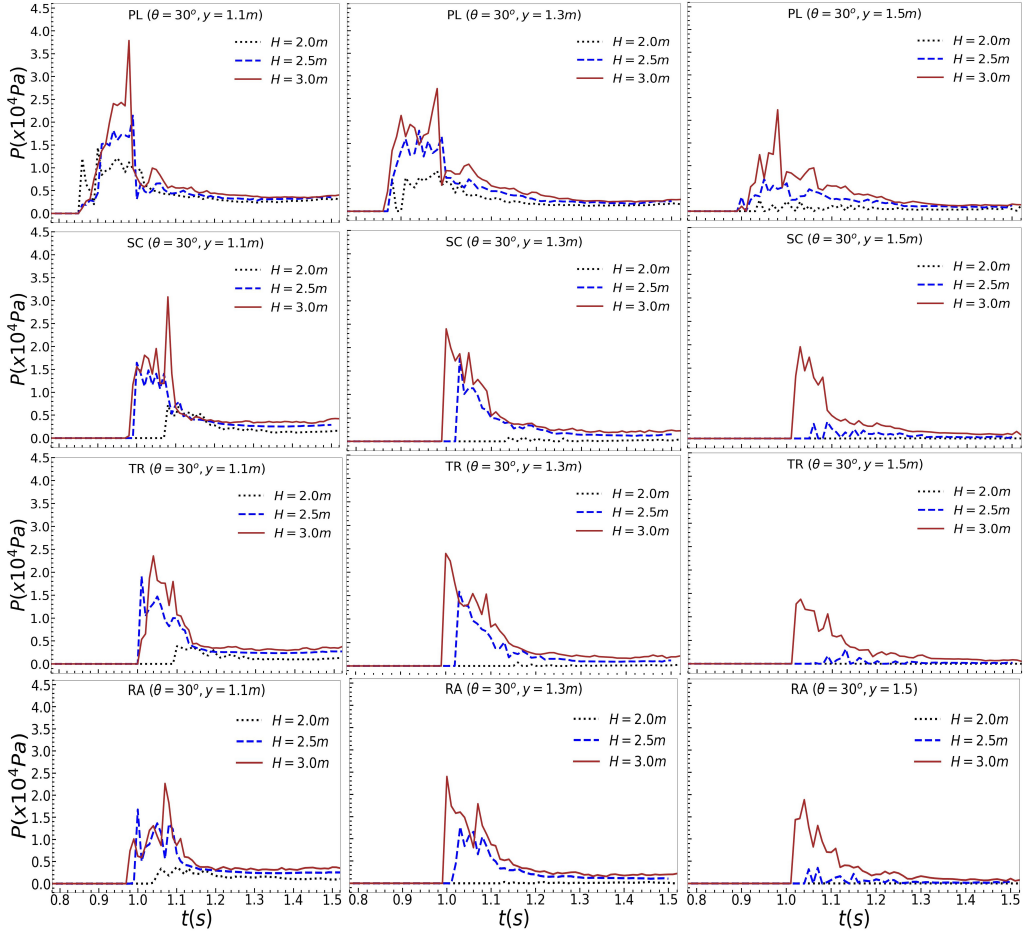}

	\caption{Variation of pressure on the wall at $y=1.1$m, $1.3$m and $1.5m$ over time for various $H$ and RSC.}
	\label{Fig:PresHieght30deg}
\end{figure} 
%
The time evolution of pressures at $y=1.1$m, $1.3$m and $1.5$m on the downstream coastal infrastructures for different $H$ and various RSCs are displayed in Fig. \ref{Fig:PresHieght30deg}. 
For PL, the wave hits the wall for all $H$ and all $y$ values considered here. 
The $P$ versus $t$ curve exhibits a peak structure except in cases where the pressures on the wall due to the water wave are insignificant (for example, the $H=2.0$m case). The peak pressure for PL is not only higher than that of other RSCs but also occurs a little earlier than the other RSCs.
We noted that a higher $H$ corresponds to a higher pressure for all RSC, as expected. The pressure on the wall due to the water wave decreases as one moves from the bottom ($y=1.1$m) to the top ($y=1.5$m). Eventually, $P$ becomes zero at a value of $y$ where the water can not reach. For $H=1.0$m and $1.5$m, the fluid does not climb the ramp wall (except for PL); hence, the wall pressure for other RSC is insignificant. A discussion on the quantitative comparison of $P^m$ is given in Table 2.

\begin{table}
	\centering	
	\caption{Maximum pressure (in Pa) at the wall for ramp of $\theta=30^\circ$. Second row give percentage of change in $p^m$ for various RSC with respect to PL}
	\begin{tabular}{|c|c|c|c|c|c|}
		\hline
		$H$(m) & $y$(m) & $P^m_{PL}$(Pa) &$P^m_{SC}$(Pa) & $P^m_{TR}$(Pa) & $P^m_{RA}$(Pa)\\
		\hline
		\multirow{5}{*}{2.0} &1.1 & 14444 & 5520 & 3855 & 3753 \\
		&   &          & (61.70\%)   & (73.31\%)   & (74.01\%)   \\
		\cline{2-6}
		 & 1.2 & 12678& 2516& 1538 & 1111 \\
		&   &          &(80.15\%)   & (87.87\%)    & (91.23\%)   \\
		\cline{2-6}
		 & 1.3 & 8913  & 979   & 753  & 317 \\
		&   &          & (89.01\%)    & (91.55\%)    & (91.27\%) \\
		\cline{2-6}
		 & 1.4 & 4168 & 1038  & 370   & 207 \\
		&   &          & (75.08\%)    & (91.12\%)    & (95.03\%) \\
		\cline{2-6}
		& 1.5 & 2468  & 380  & 29   & - \\
		&   &          & (84.60\%)     & (98.80\%)     & - \\
		\hline
		\multirow{5}{*}{2.5} & 1.1 & 25441 & 16451&	19099&	16749
		\\
		&   &          &35.33\%	&24.93\%&34.16\%
		\\
		\cline{2-6}
		& 1.2 & 25561 & 21456	&15101&	16140
		\\
		&   &          &  16.06\%&	20.21\%&35.95\%
		\\
		\cline{2-6}
		 & 1.3 & 19872 & 17705	&15856&	12728
		\\
		&   &          &   49.55\%&	64.46\%&67.52\%
		\\
	  \cline{2-6}
		& 1.4 &13253& 6686&	4710&	4304
		\\
		&   &          &49.55\%&	64.46\%&	67.52\%
		\\
  	\cline{2-6}
		& 1.5 & 7095&3573&	3120&	3510
		\\
		&   &          & 49.64\%	&56.03\%	&50.52\%
		\\
		\hline
		\multirow{5}{*}{3.0} & 1.1 & 37885& 30795 & 23551 & 22642 \\
		&   &          & (18.71\%)   & (37.84\%)    & (40.23\%)   \\
		\cline{2-6}
		& 1.2 & 40675& 37150 & 29896& 24703 \\
		&   &          & (8.67\%)     & (26.50\%)     & (39.27\%)   \\
		\cline{2-6}
		& 1.3 & 27195& 24901    & 24150    & 23877 \\
		&   &          & (8.43\%)    & (11.20\%)    & (12.20\%) \\
    	\cline{2-6}
		& 1.4 & 27145& 25000    & 25259 & 25297 \\
		&   &          & (7.90\%)      & (6.95\%)     & (6.81\%) \\
		\cline{2-6}
		& 1.5 & 22330& 19568 & 13767 & 18821 \\
		&   &          & (12.37\%)    & (38.34\%)    & (15.71\%) \\
		\hline
	\end{tabular}
	\label{PmaxVSyTh30H23}
\end{table}
%
Maximum pressure ($P^m$) for various RSCs at different values of $y$ on the wall for $H=2.0$m, $2.5$m and $3.0$m, $\theta=30^o$ are tabulated in table \ref{PmaxVSyTh30H23}. The percentage of decrease in $P^m$ for each of the RSCs in comparison to those of PL is displayed in parentheses.
We note for $H=2.0$m the $P^m$ is largest (1444 Pa) at the bottom of the wall ($y=1.1$m), which decreases with increase of height of the point ($y$) as was seen from the previous figure (Fig.\ref{Fig:PresHieght30deg}). For $H=2.5$m and $3.0$m, $P^m$ is observed at $y=1.2$m, which is a little above the bottom point. The reason is that, in this case, the water jet directly hits the wall at this point (as evident from the second and third rows of Fig. 7). 
We observe that at $H = 2.0$m, the most significant reduction in pressure is achieved by RA, followed by TR, and then SC. The maximum decrease in $P^m$ is observed for RA, which is about $91\%$ to $95\%$ ($y=1.4$m). 
As expected, the plain slab has the least effect on momentum reduction, leading to the largest pressure on the wall. A similar trend is observed for $H=2.5$m.  But for $H=3.5$m  the maximum reduction of $P^m$ is around $40\%$ ($y=1.3$m and $1.4$m). 

\begin{figure}[ht]
	\centering
	\includegraphics[width=0.7\textheight]{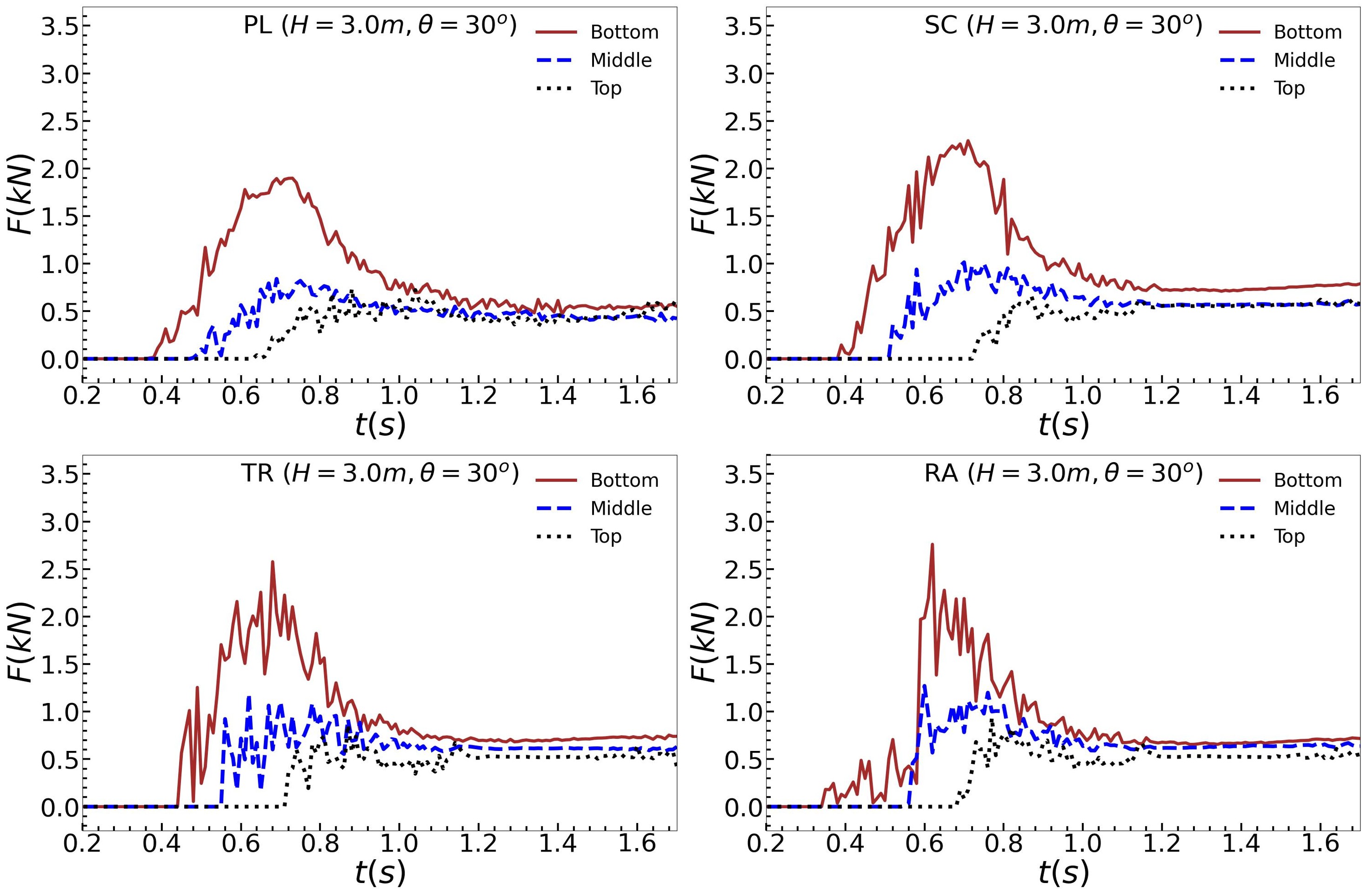}
	\caption{Net Force  acting on three sections located at bottom, middle and top of the ramp for $H=3.0$m.}
	\label{Fig:ForceValueShape}
\end{figure}
\begin{figure}
	\centering
	\includegraphics[width=0.7\textheight]{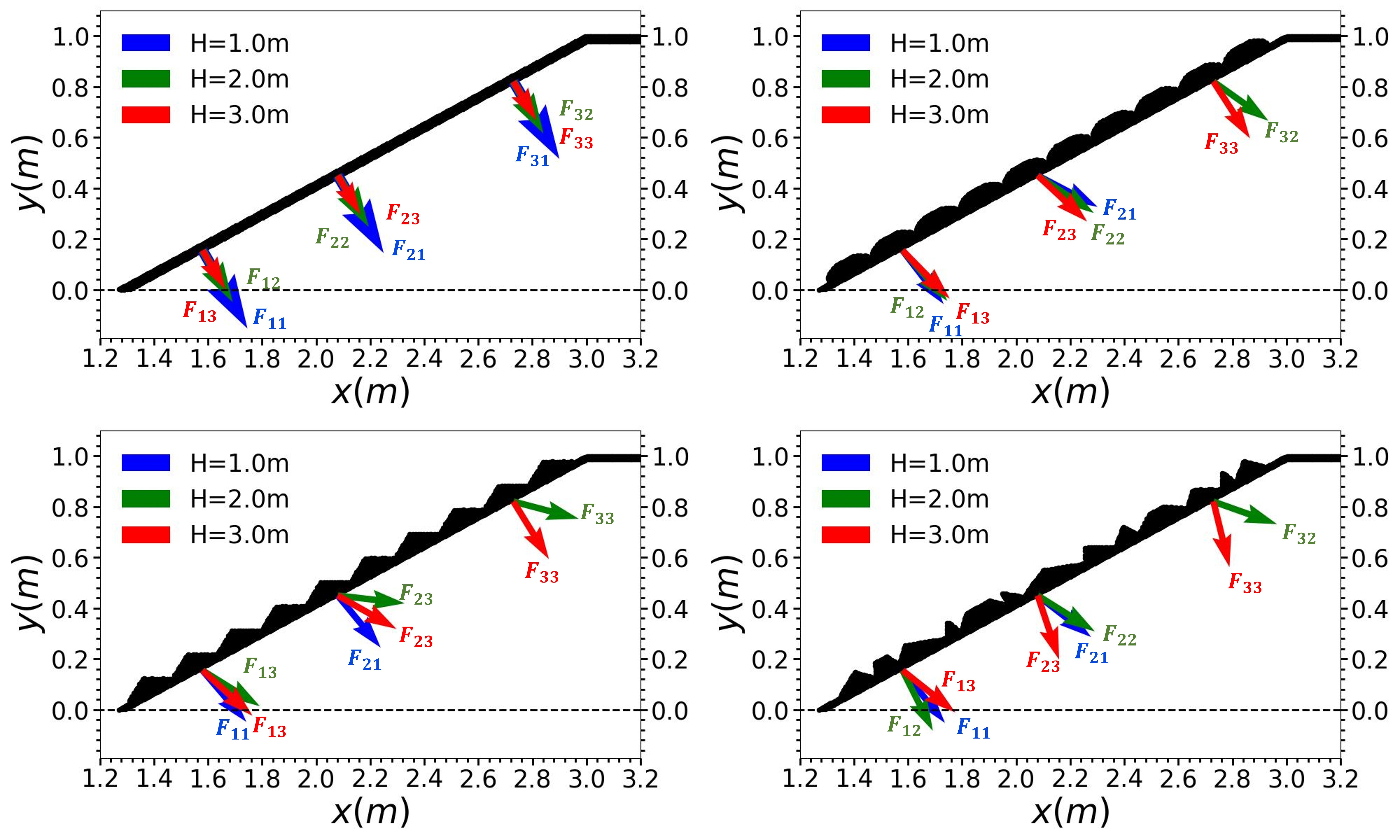}
	\caption{The direction of maximum force  on the ramp on a unit of structure at different locations for different $H$. 
}
\label{Fig:ForceDirection30}
\end{figure}

%
The net forces $F_{ij}$ on three sections of the ramp  (bottom, middle, and top)  are computed for $\theta=30^o$ and various ramp configurations, PL, SC, TR and RA. The index, $i(=1,2$ and $3)$ corresponds to the location of the section span within $dl=0.2$m (one unit structure repeated for revetment of the ramp) centered at  ($x,y$): bottom (1.062m, 0.170m), middle (1.55m, 0.460m), top (2.003m, 0.84m), whereas, $j$ ($=1,2,$ and $3$), represents dam heights of $H=1.0$m, $2.0$m, and  $3.0$m. Time evolution of net forces $F$(kN) is displayed in Fig. \ref{Fig:ForceValueShape} for $H=3.0$m and $\theta=30^o$. The directions of forces (not the magnitude) when its magnitude is the largest for various $H$ are shown in Fig. \ref{Fig:ForceDirection30}. It is observed from Fig. \ref{Fig:ForceValueShape} that the force acting on a particular RSC is maximum at the bottom section and is the least at the top section. In all cases, the $F$ versus $t$ graph shows a peak-structured pattern with fluctuations. In the bottom section, peaks are mostly around $0.7$s, except for RA, which peaks at $0.6$s. The kinetic energy of the incoming wave remains the same for a particular value of $H$. The wave loses a fraction of its energy when it hits the ramp, and the remaining energy is carried towards the wall. In this case, the maximum force exerted by the wave on the ramp with RA and the least for PL, which assures that the pressure on the wall is more for PL than other cases and a maximum reduction of $P$ on the wall occurs for RA. A comparison of numerical values is presented in table \Ref{tab:3_0}, which corroborates the observations from Fig.  \ref{Fig:ForceValueShape}.
\begin{table}[ht]
	\centering
	\caption{Maximum force experienced on sections of $30^o$ ramp.($y_1,y_2$):bottom (0.0125,0.22), middle (0.41,0.51), top (0.79,0.89)} 
	\begin{tabular}{|c|c|c|c|c|}
		\hline
		\multirow{2}{*} RSC & $H$ &\multicolumn{3}{|c|}{F(N)}\\
		\cline{3-5}
		&  & Bottom & Middle & Top \\
		\hline
		\multirow{4}{*}{PL } 
		&1.5 &622.63 & 374.34& 321.00 \\
		&2.0 &1118.35 & 499.11 &496.57  \\
		&2.5 &1666.67&681.52& 643.22  \\
		&3.0 &1898.26 & 840.30 &736.68  \\
		\hline
		\multirow{4}{*}{SC} 
		&1.5 &1164.79 & 716.33 & 324.07 \\
		&2.0 &1642.48 & 741.67 & 776.66  \\
		&2.5 &2557.96& 1256.91& 892.74 \\
		&3.0 &2292.31 & 1014.67 &659.87  \\
		\hline
		\multirow{4}{*}{TR} 
		&1.5 &913.72 &  531.11&374.32  \\
		&2.0 &1600.81 & 708.17 &661.53  \\
		&2.5 &2017.67& 1034.06&689.35 \\
		&3.0 &2576.65 &  1185.22&862.31  \\
		\hline	\multirow{4}{*}{RA} 
		&1.5 &3277.02 &  1430.49&860.31  \\
		&2.0 &2299.71 & 1461.59 &948.90  \\
		&2.5 &2365.77& 1002.75& 818.72  \\
		&3.0 &2759.95 &1271.25 & 936.39 \\
		\hline	
	\end{tabular}
	\label{tab:3_0}
\end{table}
%

\begin{figure}[ht]
	\centering
	\includegraphics[width=0.7\textheight]{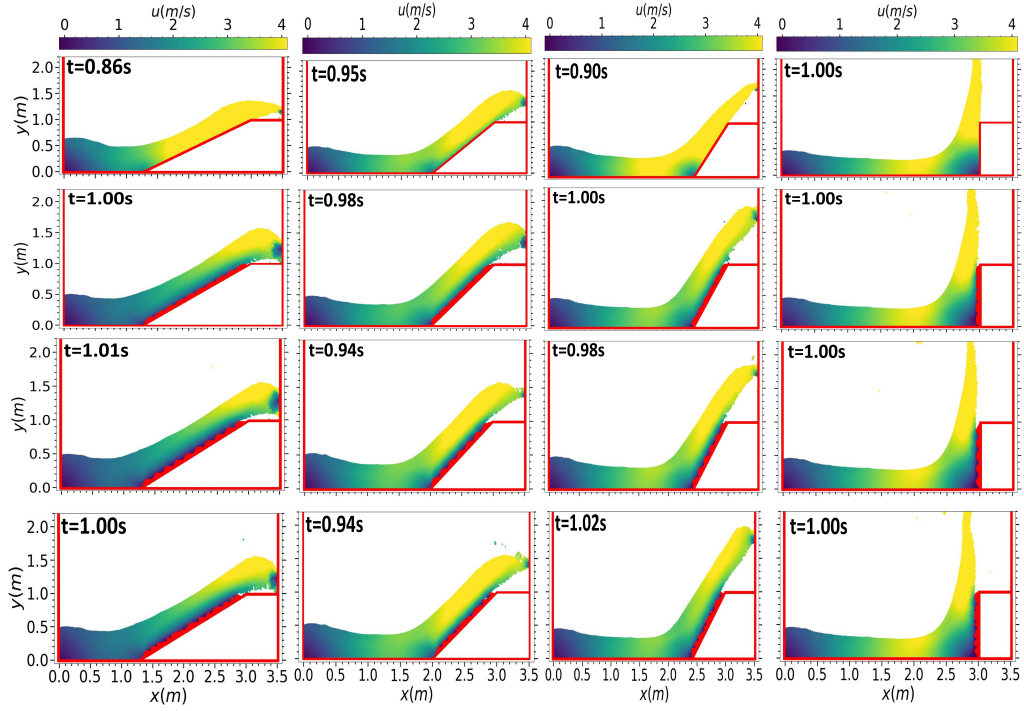}
	\caption{Velocity contour of the flow at time when the water wave impact on the wall first for for $H=3.0m$ for various $\theta$.}
	\label{Fig:dynaH3}
\end{figure}

We vary the amplitude $A_m$ (0.050m, 0.075m, and 0.100m) of the two structures on the ramp surface, SC and TR and computed the maximum pressure $P^m$ on the wall for $H=3.0$m, which are displayed in table \ref{tab:4_0}. $P^m$ are located at $y=1.2$m for all three  $A_m$s considered here. 
We note that the larger the amplitude ($A_m$) of the structures on the ramp, the greater the reduction in $P^m$, suggesting that one needs to adjust the amplitude based on the energy of the incoming wave (velocity) to mitigate damage to the shoreline and coastal structures.
\begin{table}[ht]
	\centering
	\caption{Maximum pressure on wall due to amplitude variation}
	\begin{tabular}{|c|c|c|}
		\hline
		$A_m$(m) &  $P$ (SC)  & $P$ (TR)  \\
		\hline
		0.050&37150.65 &29896.93  \\
		0.075& 25878.74 & 20327.83\\
		0.100&21420.66  &18757.66 \\
		\hline
	\end{tabular}
	\label{tab:4_0}
\end{table}
We vary $\theta$ from  $30^o$ to $90^o$ in steps of $15^o$ to analyse the effect of the inclination of the ramp on the flow impact on the coastal infrastructure. Figure \ref{Fig:dynaH3} compares the flow behaviour (velocity contour) for $H=3.0$m and various topographies of the ramp's surface. 

\noindent The first row depicts the velocity contour of flowing water at $t$ when the water jet first contacts the wall for all ramp angles except $\theta=90^o$, in which case the water jet moves up vertically and does not hit the coastal infrastructure (wall). When the slope of the ramp is increased from $30^\circ$, the fluid is striking the vertical wall at a slightly higher $y$ and the impact time is also reduced in comparison to the $30^\circ$ ramps, which is clear from the numerical values presented in table \ref{tab:imp}.
\begin{table}
	\centering
	\caption{Initial impact height $y^i$ of water on vertical wall  for the ramp angle $\theta=30^0,45^0$ and $60^0$. The time of contacting the wall first are shown the the parenthesis.} 
	\begin{tabular}{|c|c|c|c|c|c|}
		\hline
		\textbf{$\theta$} &\textbf{H(m)} & $y^i$ (PL) & $y^i$ (SC)  & $y^i$ (TR)& $y^i$ (RA)\\
		\hline
		$30^o$&3.0 & 1.15 (0.86s) & 1.20 (1.00s)&1.25 (1.01s)&1.20 (1.00s) \\
		\hline
		$45^o$&3.0 & 1.40 (0.95s) & 1.40 (0.96s)&1.45 (0.94s)&1.45 (0.94s) \\
		\hline
		$60^o$&3.0  & 1.70(0.90s)&1.80 (1.00s) &1.75 (0.98s) &1.80 (1.02s) \\
		
		\hline
	\end{tabular}
	\label{tab:imp}
\end{table}

We present the maximum pressure ($P^m$)  on the wall for  $\theta=30^o$, $45^o$ and $60^o$ for various RSCs in table \ref{tab:6a}. We consider $H=2.5$m and $3.0$m. For a particular RSC, $P^m$ decreases with increasing the angle as the component of the flow velocity $v$ along the ramp surface, is proportional to $cos \theta$ since the flow of water before impacting the ramp is mostly along the horizontal ($x$) direction. The same trend is observed for all (considered) RSC and $H$. As for the effect of ramp topography on $P^m$, we find that $P^m$ is reduced to its maximum value for RA. 
RA offers the minimum $P^m$  in all cases considered here.
\begin{table}
	\centering
	\caption{Maximum pressure and location of maximum pressure $P^m$ on the wall for various  $\theta$ and RSC for $H=2.5$m and $3.0$m. 
	}
	\begin{tabular}{|c|c|c|c|c|c|}
		\hline
		$H$(m) & $\theta$ & $P^m$ (PL) &$P^m$ (SC)& $P^m$ (TR) & $P^m$ (RA) \\
		\hline
		\multirow{3}{*}{2.5} & $30^o$ &25561.26   &21456.23 &19099.77 &16749.35  \\
		& $45^o$ &14119.55 &13092.46 &12870.30 &11431.63\\
		& $60^o$ &12968.86 &9041.38  &8914.99 & 8502.72 \\
		\hline
		\multirow{2}{*}{3.0} & $30^o$ & 40675.36 & 37150.65 & 29896.93 & 24703.16 \\
		& $45^o$ & 16716.81& 13605.02 & 12410.15 &  12136.84\\
		& $60^o$ & 15706.50  & 12409.90 & 11157.89 &  9628.88\\
		\hline
	\end{tabular}
	\label{tab:6a}
\end{table}
\begin{figure}[ht]
	\centering
	\includegraphics[width=0.7\textheight]{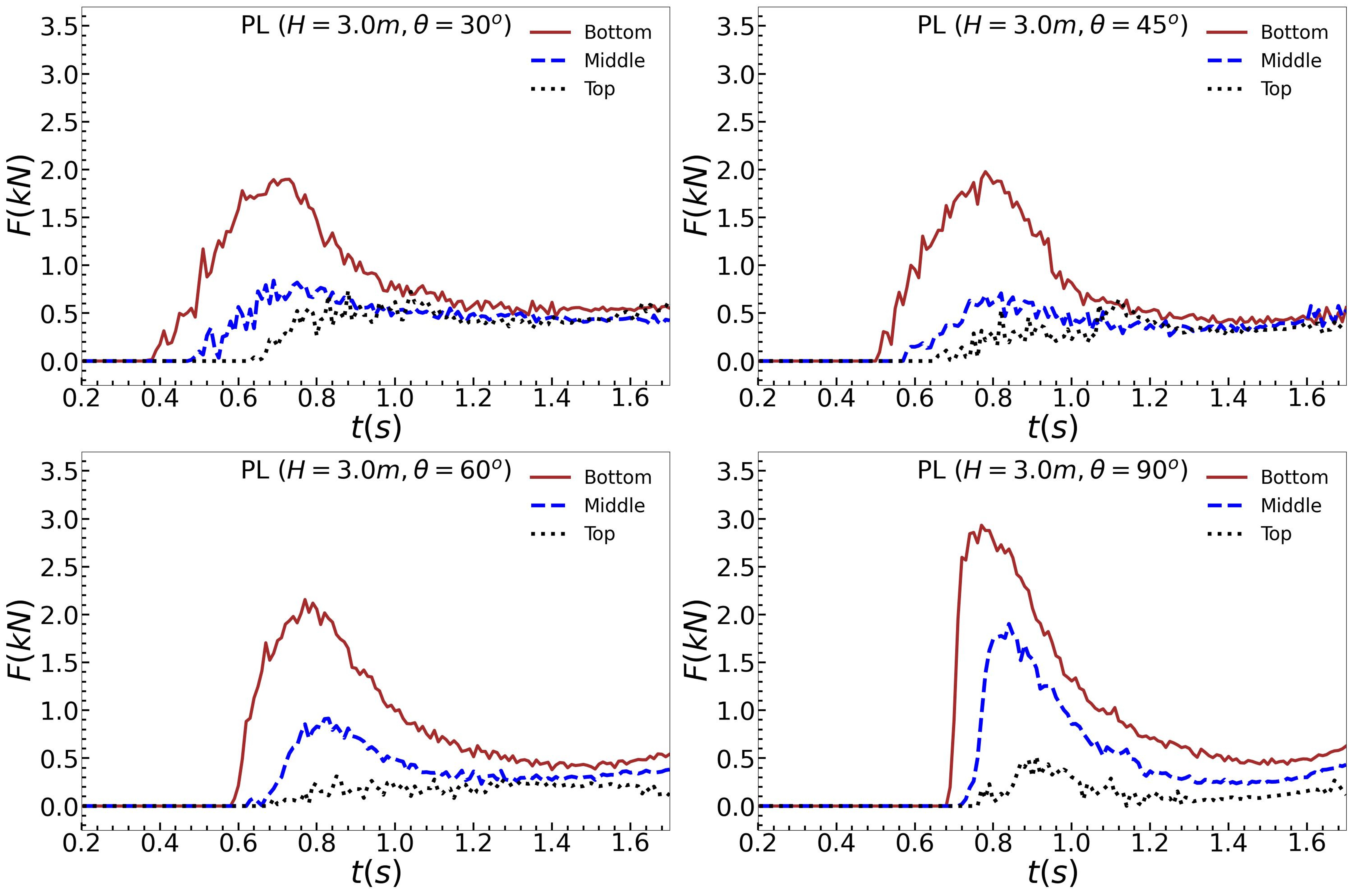}
	\caption{Net Force  acting on local structures located at bottom,middle and top of the ramp at $H=3.0m$ for plain ramp}
	\label{Fig:ForceValuePlain}
\end{figure}

We consider ramps with plain surface (PL) at angles, $\theta=30^o$, $45^o$, $60^o$ and $90^o$ to study the effects of ramp's inclination on the water wave pressure on the coastal structure. We focus on three unit sections for each of them, with a length along the ramp's surface of $dl=0.2$m, centred such a way that the water wave hits the centres of each structures at the same height, as mentioned for the corresponding structures (top, middle, and bottom) in the discussion of Fig.  \ref{Fig:ForceValueShape}. The variation of the net force $F(kN)$ due to the water wave hitting on each section with time is plotted in Fig. \ref{Fig:ForceValuePlain} for $H=3.0$m. For all cases studied here, the bottom and middle sections show peak structured curves, whereas the force on the top sections are almost flat and negligible. The bottom sections  experience the maximum force at $t \approx 0.7$s to $0.8$s, whereas the force on the top sections are negligible. For the other angles, the $P^m$ are in between those for $30^o$ and $90^o$  (Fig. \ref{Fig:ForceValuePlain}). $P^m$ on the each ramp section increases with increasing $\theta$, implying that the pressure on the wall will be reduced as $\theta$ increases. But with a larger $\theta$, the impact of the water wave on the ramp (seashore) will be increased and hence the chances damage to the ramp. 
\section{Conclusion }

We present a comprehensive analysis of the interaction of water wave (dam break model is considered as the wave-maker) with the ramp of various topographies and coastal infrastructure (wall). We studied pressure distribution on the wall, time evolution of the force on the ramp, and fluid-particle tracking in different situations. The findings highlight is that the ramp's surface structures have a significant influence on the flow behaviour, particularly in reducing pressure on the coastal infrastructures for specific dam heights. The dam-break flow is validated by comparing its results with the experimental data. 

Tracking individual fluid particles reveals distinct motion patterns based on ramp's shape, with variations in impact heights and descent behaviour.
For $\theta=30^o$, the plain ramp (PL) consistently exhibits the highest pressure on the wall, resulting in the most significant potential for structural damage. In contrast, the RA slab provides the lowest water pressure on the wall for all  $H$. At $H=2.0m$, ramp structures with an amplitude $A_m=0.05m$  effectively mitigate wall pressure, demonstrating their ability to manage the impact of the water flow. However, at $H=3.0m$, the influence of ramp structures is minimal, as the observed peak pressure is comparable to that of the plain slab. 

An increase in the amplitude of the structure on the ramp's surface for revetment, results in a larger reduction to the maximum pressure of the wall, which suggests that the amplitude of the structure on the ramp should be fixed according to the energy (velocity) of the wave in order to get sufficient reduction in $P^m$ to protect the coastal structure.

A part of energy of the incoming water wave is dissipated due to the interaction of the wave with the  the ramp. Increasing the ramp angle $\theta$, more energy is dissipated along the ramp surface, leading to a reduction in pressure on the coastal wall (structure). The calculation of the net force across the three ramp sections consistently shows that the bottom section experiences the highest force while the top section experiences the least force due to reduced direct impact. 
 
These findings underscore the importance of integrating coastal topographical data into onshore flood risk assessments and design frameworks. The study advocates for a hybrid approach combining engineered resilience with natural coastal defences to enhance the safety and longevity of onshore infrastructure in flood-prone regions.

The novelty of this work lies in its focused exploration of ramp topography as a passive energy dissipation mechanism in coastal flood scenarios, using SPH-based simulation.
Insights from this study underscore the importance of ramp geometry in dam-break flow design, offering potential ways for optimising structural resilience and energy dissipation in hydraulic engineering applications.
%

%
  \bibliographystyle{elsarticle-num-names} 
  \bibliography{ref1}



%
%
%
\end{document}